\documentclass[pre,showpacs,amsfonts,amssymb,floatfix]{revtex4-1}
\usepackage{amsmath}
\usepackage{graphicx}
\usepackage{bm}
\usepackage{color}
\allowdisplaybreaks

\begin{document}
\title{Recurrence time correlations in random walks with preferential relocation to visited places}
	\author{Daniel Campos}
\affiliation{Grup de F{\'i}sica Estad\'{i}stica.  Departament de F{\'i}sica.
	Facultat de Ci{\`e}ncies. Edifici Cc. Universitat Aut\`{o}noma de Barcelona,
	08193 Bellaterra (Barcelona) Spain}
\author{Vicen\c{c} M\'{e}ndez}
\affiliation{Grup de F{\'i}sica Estad\'{i}stica.  Departament de F{\'i}sica.
	Facultat de Ci{\`e}ncies. Edifici Cc. Universitat Aut\`{o}noma de Barcelona,
	08193 Bellaterra (Barcelona) Spain}

\begin{abstract}
Random walks with memory typically involve rules where a preference for either revisiting or avoiding those sites visited in the past are introduced somehow. Such effects have a direct consequence on the statistics of first-passage and subsequent recurrence times through a site; typically, a preference for revisiting sites is expected to result in a positive correlation between consecutive recurrence times. Here we derive a continuous-time generalization of the random walk model with preferential relocation to visited sites proposed in [\textit{Phys. Rev. Lett.} 112, 240601] to explore this effect, together with the main transport properties induced by the long-range memory. Despite the highly non-Markovian character of the process, our analytical treatment allows us to (i) observe the existence of an asymptotic logarithmic (ultraslow) growth for the mean square displacement, in accordance to the results found for the original discrete-time model, and (ii) confirm the existence of positive correlations between first-passage and subsequent recurrence times. This analysis is completed with a comprehensive numerical study which reveals, among other results, that these correlations between first-passage and recurrence times also exhibit clear signatures of the ultraslow dynamics present in the process.
\end{abstract}

\maketitle

\section{Introduction}

Purposeful memory is one of the essential ingredients that serves us to distinguish the behavior of higher organisms from that of simpler (living or non-living) entities; this is an idea which has been recognized for very long and has represented a matter of debate in science, psychology and epistemology for decades. Nowadays, quantitative approaches to animal and human behavior have progressively become a field of active research in biological physics. The study of neuronal patterns in the brain with the help of imaging techniques and its modelization through mathematical tools from network theory and/or population dynamics represent a prime example \cite{chialvo10,papo17}, but many other could be cited.

In particular, for those behavioral processes related to dispersal and/or navigation of living beings, it is clear that random walks represent a convenient description in order to condense and/or account for the properties of real trajectories and individual/collective space-use \cite{codling08,mendez14,bartumeus16}. Simpler approaches within this area just neglect any effects from spatial memory, so the position of the individual walker $X(t)$ is then a Markovian random variable. However, a rich bibliography on random-walks with memory rules of different sort has been developed throughout the years and connected to biological movement, too. Some examples of this include models in which walkers tend to exhibit locally some kind of directional memory or persistence (these could be loosely termed as \textit{annealed} memory models); this is the case of Persistent random walks \cite{holmes93,selmeczi05,codling08}, Continuous-Time random walks (CTRW) \cite{shlesinger18}, L\'{e}vy Walks \cite{viswanathan96,ariel15}, etc. On the other side, \textit{quenched} memory models represent a more complex situation in which local information about the sites visited is stored by the walker somehow, and so future rules of advance will explicitly depend on it, which in general makes the process highly non-Markovian. Some well-known frameworks falling within this class are the different versions and generalizations of the self-avoiding random walk (though originally this model was proposed to describe polymer growth, not biological movement) \cite{freund92}, elephant random walks (which were probably the first class of solvable models proposed with long-range memory effects) \cite{schutz04}, or, more recently, models with preferential relocation to already visited sites \cite{boyer14,falcon17} or preferential persistence for familiar paths \cite{abramson14,kazimierski15}.

Though a large amount of realism can be gained by introducing memory in the description of animal and human movement, it is clear that mathematical treatment becomes then cumbersome due to its non-Markovian character. First-passage and coverage properties, for instance, of non-Markovian processes represent a formidable problem for which it is very difficult to extract analytical results unless additional assumptions are considered \cite{guerin16}. For the case of \textit{annealed} memory some exceptions can be found, like those works where non-stationary random walk patterns are considered \cite{campos16}, but for \textit{quenched} memory it is very difficult to find references in the literature where this has been even addressed; we can cite the recent work by Kearney and Martin \cite{kearney16} on the first-passage properties of P\'{o}lya urns and their connection to random walks as one of the few exceptions.

Taking all this context into account, our aim is to propose the study of first-passage and recurrence times statistics for \textit{quenched} memory random walks roughly describing the movement of biological organisms with different memory (cognitive) abilities. A whole analytical treatment of such situations, as mentioned above, is in general unattainable but still some general properties of interest can be derived in many situations numerically. If $\Theta_n$ represents a random variable describing the time at which the individual hits a site for $n$-th time, then we will focus here on the random variables
\begin{eqnarray}
\nonumber T_1 &=& \Theta_1 \\
\nonumber T_n &=& \Theta_n-\Theta_{n-1}, \quad n>1
\end{eqnarray} 
representing the random times between consecutive hittings. So, a consequence of introducing memory effects in the trajectories could be the emergence of correlations between $T_1,T_2,\ldots$. Note that all these variables would be independent for the simplest case of Markovian random walks, since memory is lost after hittings. So, characterizing the correlations between these recurrence times could provide a way to classify memory random walks.

As a first step towards this aim, we here study the correlation properties between first-passage and successive recurrence times in a generalized (continuous-time) version of the random-walk with preferential relocation to visited sites \cite{boyer14,boyer14b}. This preferential relocation process, as reported in the original work, tries to capture some basic properties of foraging in higher animals driven by a tendency to revisit sites where resources (e.g. food,...) have been successfully detected previously (so assuming that these resources are never depleted and/or can be replenished in a relatively short time). Hence, this process is obviously expected to yield a positive correlation between the variables $T_n$, though this idea has never been explored previously as far as we know.

The present article is structured as follows. In Section \ref{model} we derive the Continuous-Time Random Walk master equation for random walks with preferential relocation to visited sites, and justify the interest of such generalization if compared to the original (time-discrete) version. In Section \ref{dispersal}, we study the dispersal properties through the Mean Square Displacement (MSD) of the corresponding random walkers to check that our results are in agreement with those from the original model. In section \ref{passage} we formally present the hitting and recurrence problem for this case and provide analytical justification to support the existence of positive correlations between first-passages and sussequent recurrence times for a particularly simplified case. Also, we show results from Monte Carlo simulations in order to understand the main properties of the process. Finally, the conclusions from our study are presented in Section \ref{conclusions}.

\section{Continuous-Time model} \label{model}

\subsection{CTRW framework}

We start by revising briefly the classical CTRW formalism to facilitate understanding of the model presented below. The CTRW is based on the idea that the walker performs jumps of random (i.i.d.) lengths separated by random (i.i.d.) waiting times. A possible mathematical derivation is based on the combination of two balance equations. The first one states that the probability of reaching a position $x$ at the $m$-th step, denoted by $j_m(x)$, satisfies the mesoscopic balance equation
\begin{equation}
j_m(x)= \int_{-\infty}^{\infty} j_{m-1}(x-z) \Phi(z) dz + \delta_{m,0} \delta (t) j_0,
\end{equation}
where $\Phi(x)$ is the jump kernel, which determines the probability distribution function of the jump lengths, and the last term stands for the initial condition (where $\delta_{m,0}$ is a Kronecker Delta function, $\delta (t)$ is a Dirac Delta function, and $j_0=\delta(x-x_0)$ with $x_0$ the position of the site occupied at $t=0$).
One can also include time explicitly within this expression to write
\begin{eqnarray}
j_m(x,t)&=& \int_{0}^{t} \int_{-\infty}^{\infty} j_{m-1}(x-z,t-\tau) \varphi(\tau) \Phi(z) dz d\tau \nonumber \\
&+& \delta_{m,0} \delta (t) j_0,
\label{eq1}
\end{eqnarray}
where $j_m(x,t)$ corresponds to the probability that the $m$-th step is done to position $x$ at time $t$ (so $j_m(x)=\int_{0}^{\infty} j_m(x,t)dt$) and $\varphi(t)$ is the probability distribution function of waiting times between consecutive jumps.

A second equation is introduced through the probability of being at position $x$ at time $t$ after $m$ steps have been made, $p_m(x,t)$; this expression reads
\begin{equation}
p_m(x,t) = \int_0^{t} j_m(x,t-\tau) \phi(\tau) d\tau ,
\label{eq2}
\end{equation}
where $\phi(\tau)$ is the probability that the walker has not jumped in a time $\tau$ since it arrived to $x$, so this satisfies $\phi(t)=\int_{t}^{\infty} \varphi(\tau) d \tau$. So, by combining (\ref{eq1}) and (\ref{eq2}) one obtains the master equation which contains the statistical properties of the CTRW. In particular, if we sum the equations (\ref{eq1}) and (\ref{eq2}) for $m$ from $0$ to $\infty $ (with $j_m=0$ for $m<0$) we recover the well-known master equation of the CTRW as it appears in the books on the subject \cite{mendez14}:
\begin{equation}
j(x,t)= \int_{0}^{t} \int_{-\infty}^{\infty} j(x-z,t-\tau) \varphi(\tau) \Phi(z) dz d\tau + \delta (t) j_0
\end{equation}
\begin{equation}
p(x,t) = \int_0^{t} j(x,t-\tau) \phi(\tau) d\tau 
\label{prop}
\end{equation}

\subsection{Preferential relocation to visited sites}
\label{preferential}

Now we include in the CTRW framework the possibility that the walker can use memory to return to previously visited sites.  While the original model with preferential relocation rules was discrete both in time and space \cite{boyer14}, a continuous generalization has already been proposed in a different context \cite{boyer17}, though a mesoscopic derivation was not provided there as we do in the following. In particular, a generalization within the CTRW framework allows one the possibility to consider in a natural way different waiting time distributions, including those with power-law tails (leading to a L\'{e}vy statistics) or with a combination of different dispersal modes, for example.

We will consider that at the end of each waiting time the particle can either (i) decide to do a random jump governed by the kernel $\Phi(x)$ with probability $\alpha$ (in the following, we denote this as the \textit{normal} transport mode), or (ii) use its memory and then \textit{fly} instantaneously to a previously visited site with probability $1-\alpha$ (in the following, this is termed as \textit{memory} transport mode). 
If we denote the memory kernel (it is, the probability to remember a site visited $i$ jumps ago) as $K_m(i)$, with $\sum_{i=1}^{m} K_m(i) =1$ for any $m$, then we can write
\begin{equation}
j_m(x,t)= \alpha \int_{0}^{t} \int_{-\infty}^{\infty} j_{m-1}(x-z,t-\tau) \varphi(\tau) \Phi(z) dz d\tau + (1-\alpha) \sum_{i=1}^{m} K_m(i) \left[ j_{m-i}(x,t) \ast \varphi(t) ^{\ast i} \right] ,
\end{equation}
where the asterisk symbol $\ast$ denotes the time-convolution operator, and $\varphi(t) ^{\ast i}$ denotes the time-convolution of the distribution $\varphi(t)$ with itself $i$ times.

From now on, for the sake of simplicity we will use a memory kernel that gives the same weight to all previously visited sites, so $K_n(i)=1/m$, in agreement with the original model \cite{boyer14}. For that case, the previous equation becomes
\begin{equation}
j_m(x,t)= \alpha \int_{0}^{t} \int_{-\infty}^{\infty} j_{m-1}(x-z,t-\tau) \varphi(\tau) \Phi(z) dz d\tau + \frac{1-\alpha}{m} \sum_{i=1}^{m} \left[ j_{m-i}(x,t) \ast \varphi(t) ^{\ast i} \right] .
\end{equation}

We can apply Fourier-Laplace transforms to the previous equation to take advantage of the renewal property of the process in time and space,
\begin{equation}
j_m(k,s)= \alpha j_{m-1}(k,s) \varphi(s) \Phi(k) + \frac{1-\alpha}{m} \sum_{i=1}^{m} j_{m-i}(k,s) \varphi(s)^{i}.
\label{jnks}
\end{equation}
Note that we use $k$ and $s$ as the Fourier and Laplace arguments. So that, $j_m(k,s)$ represents the Fourier-Laplace transform of $j_m(x,t)$, and we will distinguish them just by explicitly writing their variables.
Now, due to the explicit dependence of the memory factor on $m$, it is also convenient to carry out a Z-transform on the jump index $m$ such that 
$$\hat{j}(\lambda,k,s)=\sum_{m=0}^{\infty} \lambda^{m} j_m(k,s),$$
with $0<\lambda <1$; by applying this on Eq. (\ref{jnks}) we obtain
\begin{equation}
\hat{j}(\lambda,k,s)= \alpha \lambda \hat{j}(\lambda,k,s) \varphi(s) \Phi(k) + (1-\alpha) \int_{0}^{\lambda} \frac{ \varphi(s) \hat{j}(u,k,s)}{1-u \varphi(s)} du.
\label{jlks}
\end{equation}

Due to the integral that appears now in the equation one needs to rewrite this expression as a differential equation by differentiating with respect to $\lambda$ and replacing the integral in the resulting equation with the help of Eq. (\ref{jlks}). This procedure leads to
\begin{equation}
\frac{d\hat{j}(\lambda,k,s)}{d\lambda} = \hat{j}(\lambda,k,s)\left[ \frac{\alpha \varphi(s) \Phi(k)}{1-\alpha \lambda \varphi(s) \Phi(k)} +  \frac{(1-\alpha)  \varphi(s)}{(1-\lambda \varphi(s)) (1-\alpha \lambda \varphi(s) \Phi(k))} \right] .
\end{equation}
The solution of this first-order differential equation reads
\begin{equation}
\hat{j}(\lambda,k,s)= \left[ \frac{1}{1-\alpha \lambda \varphi(s) \Phi(k) } \right]^{\frac{\alpha[1-\Phi(k)]}{1-\alpha \Phi(k)}}  \left[ \frac{1}{1- \lambda \varphi(s)} \right] ^{\frac{1-\alpha}{1-\alpha \Phi(k)}},
\label{jlks2}  
\end{equation}
where we have taken into account the boundary condition $\hat{j}(\lambda \rightarrow 0^{+},k,s)=1$.
Finally, taking into account that we are interested in the behavior of $j(k,s)$ (independent of $m$) we can remove the dependence on the jump index by using $j(k,s) \equiv \hat{j} (\lambda=1,k,s)=\sum_{m=0}^{\infty} j_m(k,s) $. On setting $\lambda=1$ into Eq. (\ref{jlks2}) we have
\begin{equation}
j(k,s)=  \left[\frac{1}{1-\alpha \varphi(s) \Phi(k) } \right]^{\frac{\alpha[1-\Phi(k)]}{1-\alpha \Phi(k)}}  \left[\frac{1}{1- \varphi(s)}\right] ^{\frac{1-\alpha}{1-\alpha \Phi(k)}}. 
\label{jks}
\end{equation}
This expression, together with (\ref{prop}) (which is still valid for the model with memory), represent the CTRW generalization of the random walk with preferential relocation to visited places. Next, we will use this result to derive several properties of the model.

\section{Dispersal properties}
\label{dispersal}

First we will explore the dispersal properties of the model in order to check that they agree with those found for the time-discrete version \cite{boyer14}. For this, we will compute the MSD $\langle x^2(t)\rangle$, which is nothing but the second moment of $p(x,t)$ in space. So that, working again in Fourier-Laplace space

\begin{equation}
\langle x^2(s)\rangle =-\lim _{k\rightarrow 0}\frac{\partial ^2 p(k,s)}{\partial k^2}=-\frac{1-\varphi(s)}{s}\lim _{k\rightarrow 0}\frac{\partial ^2 j(k,s)}{\partial k^2},
\label{msd1}
\end{equation}
where we have made use of Eq. (\ref{prop}). Performing the second derivative of $j(k,s)$ we find, after some tedious calculations,

\begin{equation}
\langle x^2(s)\rangle =- \frac{\alpha}{1- \alpha} \frac{\Phi''(0)}{s}  \ln \left[\frac{1-\alpha\varphi(s))}{1-  \varphi(s)} \right],
\end{equation}
where $\Phi''(0)$ stands for the second derivative of $\Phi(k)$ evaluated at $k=0$. If the waiting time distribution has finite moments then we can make use of the expansion $\varphi(s) \simeq 1-\tau s+...$ for large times, with $\tau$ the mean waiting time. Hence, using this expansion and assuming $s\rightarrow 0$ we obtain
\begin{equation}
\langle x^2(s)\rangle =\frac{\alpha}{1- \alpha} \Phi''(0)\frac{\ln (\tau s)}{s},
\label{msdf}  
\end{equation}
which yields
\begin{equation}
\langle x^2(t)\rangle =-\frac{\alpha}{1- \alpha} \Phi''(0)\ln (t/\tau)\quad \text{for}\quad t\rightarrow \infty
\label{msdf2}  
\end{equation}
after inverting by Laplace. This result is very general and holds for any dispersal kernel and waiting time distributions with finite moments and predicts a ultra-slow diffusion due to the logarithmic growth of the MSD \cite{MeIo,iomin16}. If for example the dispersal kernel is Gaussian $\Phi(x)=(2\pi\sigma^2)^{-1/2}\exp (-x^2/2\sigma ^2)$
the asymptotic form of the MSD is given by

\begin{equation}
\langle x^2(t)\rangle =\frac{\alpha}{1- \alpha} \sigma ^2\ln (t/\tau)\quad \text{for}\quad t\rightarrow \infty.
\label{msdf3}  
\end{equation}

\begin{figure}[htbp]
	\includegraphics[scale=1.0]{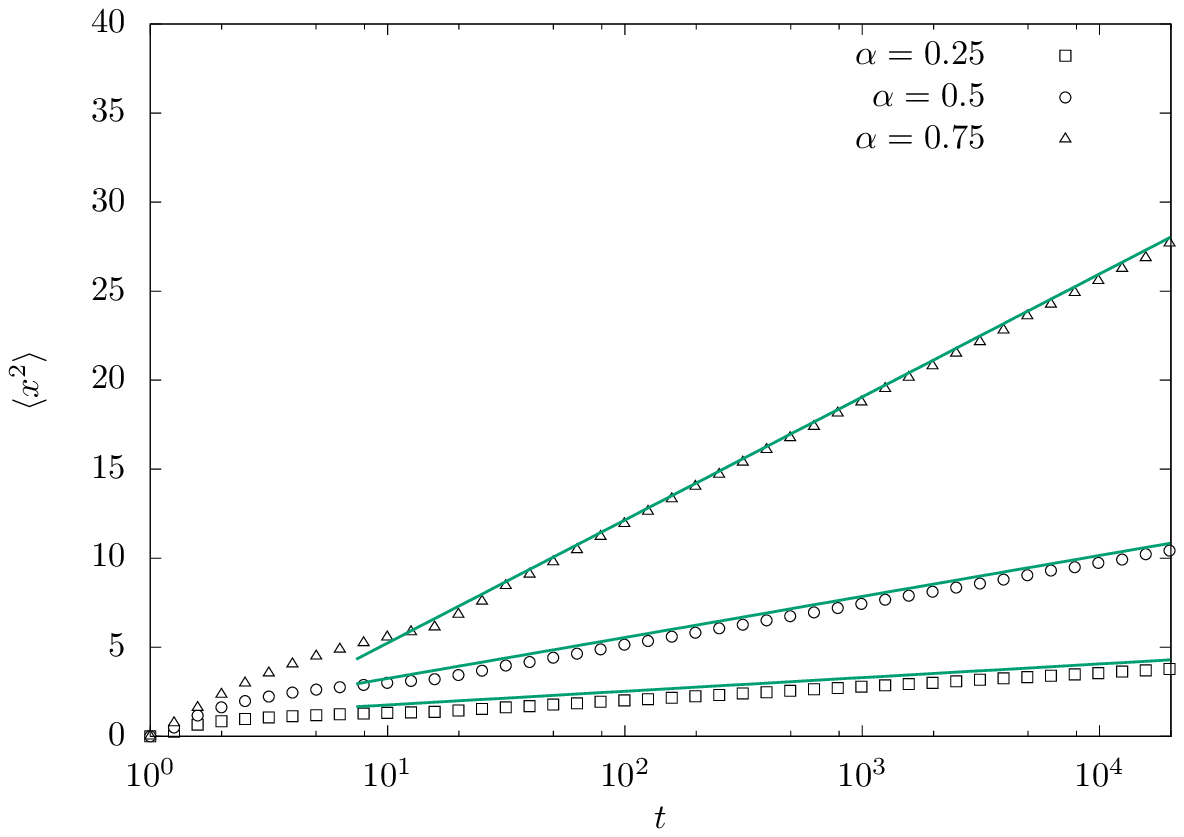}
	\caption{MSD for the CTRW model with preferential relocation to visited places. Symbols represent the results from Monte Carlo simulations averaged over $10^5$ realizations, while lines correspond to the asymptotic prediction for a Gaussian kernel from the model in Eq. (\ref{msdf3}) (appropriately shifted to facilitate visualization). For the waiting time distribution $\varphi(t)$ an exponential distribution with $\tau =1$ has been used.}
	\label{fig1}
\end{figure}

In Figure \ref{fig1} we confirm this result via direct comparison with Monte Carlo simulations of the random walk process described in Section \ref{preferential} (points) for different $\alpha$ values. We find that not only the ultra-slow (logarithmic) character of the dispersal is observed, but the prefactor $\alpha \sigma^2 / (1-\alpha)$ predicted in (\ref{msdf}) (governing the slope of the solid lines in the plot) fits almost perfectly the behavior of the simulated process. This serves as a checking of the validity of our approach prior to the analysis of the recurrence statistics we carry out in the next Section.

\section{Recurrence statistics}
\label{passage}

\subsection{Theoretical framework}
\label{theory}

First-passage times and the statistics of recurrence times for Markovian random walks can be derived in general for finite domains with the help of renewal properties \cite{condamin05}, but this no longer holds for non-Markovian walks \cite{guerin16}. While the problem is analytically unattainable for the model with preferential relocation presented in Section \ref{model}, some results can still be derived by simplifying the model considerably. 

In particular, we will use the assumption that the system is finite (with $N$ denoting the number of accessible sites) and it is spatially unconstrained (this is, the kernel $\Phi(x)$ is such that any site is accessible from any other with the same probability $1/N$). This is not unreasonable in the context of the preferential relocation model we are considering here: since the \textit{memory} mode can lead the individual to any visited site without restrictions, then it may be plausible to consider that the \textit{normal} transport mode has this spatial capacity, too. In Section \ref{numerical}, however, we will relax this assumption to check how our numerical results change when only jumps of a given maximum length are permitted.

\subsubsection{Case $n=1$} 
To illustrate our method, we first derive a formal expression for the first-passage distribution, following a similar strategy to that in Section \ref{preferential} based on discretizing the process according to the jump index $m$. We denote by $S_{1}(t)$ the survival probability through a target site $x^{*}$ (this is, the probability that after time $t$ that site has not been visited yet; note that we do not consider an explicit dependence on $x^{*}$ due to the assumption of unconstrained space). Then, this probability can be decomposed through

\begin{equation}
S_{1}(t) = \sum_{m=0}^{\infty} S_{1}^{(m)} P(m \vert t) = \sum_{m=0}^{\infty} S_{1}^{(m)} \left[ \varphi(t) ^{\ast m} \ast \phi(t) \right] ,
\label{fpt1}
\end{equation}
where $S_{1}^{(m)}$ is the survival probability after $m$ jumps, and $P(m \vert t)$ is the probability of having performed $m$ jumps at time $t$. The latter is given by the convolution of $m$ times the waiting time distribution (plus the convolution with $\phi(t)$, which is necessary to assert that the $(m+1)$-th jump has not been done yet).

Since the \textit{memory} transport mode does not contribute to the first-passage time (as the target site has not been visited yet), the survival probability after $m$ jumps is
\begin{equation}
S_{1}^{(m)}=(1- \alpha /N)^{m},
\label{fpt1b}
\end{equation}
making use of the unconstrained space assumption. This result follows since the probability to find the target at any jump is just $\alpha /N$. Transforming (\ref{fpt1}) to the Laplace space one finds
\begin{equation}
S_{1}(s) = \frac{\phi(s)}{1- \left( 1- \alpha /N \right) \varphi(s)}.
\label{fp2}
\end{equation}
So that, one can write as usual the first-passage distribution $f_1(t)$ as the time derivative of $S_{1}(t)$ or, alternatively, $f_1(s)=sS_{1}(s)-1$. The first moment of $f_1(t)$, $\langle T_1 \rangle = \int_0^{\infty} t f_1(t) \text{d}t$, provides the mean first-passage time, which reads then
\begin{equation}
\langle T_1 \rangle = \lim_{s \rightarrow 0} \frac{\text{d}f_1(s)}{\text{d}s} = \lim_{s \rightarrow 0} \left[ s \frac{\text{d}S_{1}(s)}{\text{d}s} + S_{1}(s) \right]
\label{t1}
\end{equation}
So, if we assume again that $\varphi(t)$ has finite moments (with $\varphi(s) \approx 1-\tau s+ \ldots$), then we reach from (\ref{fp2}) and (\ref{t1})
\begin{equation}
\langle T_1 \rangle = \frac{N \tau}{\alpha}
\label{t1b}
\end{equation}
and, similarly, for the second order moment
\begin{equation}
\langle T_1^2 \rangle = 2 \left( \frac{N \tau}{\alpha} \right) ^2.
\label{secondt1}
\end{equation}
These results are to be expected. In the absence of contributions from the \textit{memory} mode the process is Markovian, so (\ref{t1b}) simply represents the Wald's identity \cite{redner01} for a stochastic process with constant rate $\alpha /\tau N$ (which is nothing but the rate at which the target site will be reached through the \textit{normal} mode).

\subsubsection{Case $n > 1$} 
Now we can extend the procedure above to study subsequent recurrence times ($T_2,T_3,\ldots$, as we defined them in the Introduction) through a site. For this, we introduce the distribution function $f_n(t)$ for the random variable $T_n$ and, equivalently, the joint distribution $f_n(t;m_{n-1})$ for $T_n$ and the random variable $m_{n-1}$, which is the number of jumps done by the individual (counting since $t=0$) when it hits the site for the $(n-1)$-th time (this is, the number of jumps performed when $t=\Theta_{n-1}$). Then we can write
\begin{eqnarray}
\nonumber \langle T_n \rangle &=& \sum_{m_{n-1}=0}^{\infty} \int_{0}^{\infty} t f_n (t;m_{n-1}) \text{d}t \nonumber \\
&=& 
\sum_{m_{n-1}=0}^{\infty} \int_{0}^{\infty} t f_n (t \vert m_{n-1}) q_{n-1}(m_{n-1}) \text{d}t \nonumber \\
&=&-\sum_{m_{n-1}=0}^{\infty} q_{n-1}(m_{n-1}) \int_{0}^{\infty} t \frac{\text{d} S_n (t \vert m_{n-1})}{\text{d}t} \text{d}t,  \nonumber \\
\label{npt}
\end{eqnarray} 
where in the second step we have introduced, through the Bayes theorem, the conditional probability $f_n (t \vert m_{n-1})$ and the probability distribution function $q_{n-1}(m_{n-1})$ for the variable $m_{n-1}$. Finally, in the last step we have rewritten the conditional distribution $f_n (t \vert m_{n-1})$ in terms of the conditional survival probability $S_n (t \vert m_{n-1})$.
The reason for writing the mean recurrence time in that way is because for each recurrence process one can provide explicit expressions equivalent to (\ref{fpt1}) and (\ref{fpt1b}) for the first-passage, respectively:

\begin{equation}
S_n(t \vert m_{n-1}) = \sum_{m=0}^{\infty} S_{n \vert m_{n-1}}^{(m)} P(m \vert t) = \sum_{m=0}^{\infty} S_{n \vert m_{n-1}}^{(m)} \left[ \varphi(t) ^{\ast m} \ast \phi(t) \right]
\label{npt1}
\end{equation}
and
\begin{equation}
S_{n \vert m_{n-1}}^{(m)} = \prod_{i=0}^{m-1} \left( 1- \frac{\alpha}{N} - \frac{(1- \alpha) (n-1)}{m_{n-1}+i} \right).
\label{npt1b}
\end{equation}

Note first of all that both expressions for the survival probabilities reduce to (\ref{fpt1}) and (\ref{fpt1b}) for $n=1$ (since $m_0=0$ and so the conditional probability is unnecessary in that case). Now, we observe that the \textit{memory} mode explicitly contributes to the survival probability through the last term within the parenthesis of (\ref{npt1b}). So, the probability that the \textit{memory} mode leads the individual to the target is given by $(n-1)/(m_{n-1}+i)$, where $(m_{n-1}+i)$ is the total number of jumps performed up to date (it is, those done up to the $(n-1)$-th hitting, $m_{n-1}$, plus those done afterwards, $i$).

As a whole, the expressions (\ref{npt}-\ref{npt1b}) provide a recurrent method to determine the statistics of recurrence times as follows. First, once we know the properties of the first-passage time, we can use them to determine $q_1(m_1)$, which by definition satisfies
\begin{equation}
q_1(m_1)=S_1^{(m_1)}-S_1^{(m_1-1)}.
\label{q1}
\end{equation} 
This, in combination with (\ref{npt}-\ref{npt1b}) for $n=2$ will be used to determine $f_2 (t;m_{1})$ and its mean value $\langle T_2 \rangle$. Then we will be able to determine $q_2(m_2)$ from
\begin{equation}
q_n(m_n)= \sum_{m=0}^{m_n-1} q_{n-1}(m) \left[ S_{n \vert m}^{(m_n-m)} - S_{n \vert m}^{(m_n-m-1)} \right].
\label{qn}
\end{equation}
and the same idea can be applied recurrently for $n=3,4,\ldots$

The previous method, obviously, will become increasingly cumbersome as $n$ increases, so at practice we can only expect it to be of practical utility for $n$ small. In the next subsection we will illustrate its use for $n=2$, which is enough for the specific objectives we pursue in this paper.

\subsection{Recurrence time for $n=2$}
\label{approx}

Although the method described above could be applied to any waiting time distribution, to keep notation and results manageable we will focus here on the case of exponential waiting times, $\varphi(t)= \tau ^{-1} e^{-t / \tau}$. Note that in this case the random walk would become Markovian in absence of the preferential relocation rule (this is, for $\alpha=1$).
We already know that the mean first-passage time is determined by (\ref{t1b}). Also, introducing (\ref{fpt1b}) into (\ref{q1}) one has
\begin{equation}
q_1(m_1)= \frac{\alpha}{N} \left( 1- \frac{\alpha}{N} \right) ^{m_1-1}
\label{n21}
\end{equation}

On the other side, the combination of Eqs. (\ref{npt1}) and (\ref{npt1b}) yields for this case
\begin{equation}
S_2(t \vert m_{1}) = \sum_{m=0}^{\infty} B_{m,m_1} \left( \frac{t}{\tau} \right) ^{m} \frac{ e^{-t/ \tau}}{\Gamma(m+1)},
\label{n22}
\end{equation}
where we have defined
\begin{equation}
B_{m,m_1} \equiv \prod_{i=0}^{m-1} \left( 1- \frac{\alpha}{N} - \frac{1- \alpha }{m_{1}+i} \right).
\label{n23}
\end{equation}
So, we finally insert (\ref{n21}-\ref{n22}) into (\ref{npt}) to determine the mean recurrence time after first hitting, $T_2$. By doing this and performing the integral in $t$ we get
\begin{equation}
\langle T_2 \rangle = \sum_{m_{1}=0}^{\infty} \frac{\alpha \tau}{N} \left( 1- \frac{\alpha}{N} \right) ^{m_1-1} \sum_{m=0}^{\infty} B_{m,m_1}.
\label{t2}
\end{equation}

This expression cannot be further simplified due to the product within the definition of $B_{m,m_1}$, but we can reach useful approximations in the large-domain limit $N \gg 1$, so the number of jumps required for the first-passage, $m_1$, is also large in average. Using such approximation, one can provide an analytical approximation for $\langle T_2 \rangle$ which can be expressed as a combination of exponential integral (or, alternatively, confluent hypergeometric) functions (see the discussion in the Appendix for the details).

Also, note that the same idea can be applied to higher order or cross moments (e.g. $\langle T_2^{2} \rangle$ or $\langle T_1 T_2 \rangle$) (see Appendix). This allows us to provide an estimation for the correlation coefficient between first-passage and subsequent recurrence times
\begin{equation}
\text{Cor} (T_1,T_2) \equiv \frac{\langle T_1 T_2 \rangle - \langle T_1 \rangle \langle T_2 \rangle}{ \sigma_1 \sigma_2},
\label{cor}
\end{equation}
where $\sigma_i$ denotes the standard deviation of $T_i$. The goodness of that estimation will be checked through the comparison with results from Monte Carlo simulations in the next section.

\subsection{Numerical results}
\label{numerical}

The first results we show from our Monte Carlo simulations try to verify the validity of the approximations carried out in the previous Section. 

\begin{figure}
	\includegraphics[scale=1.0]{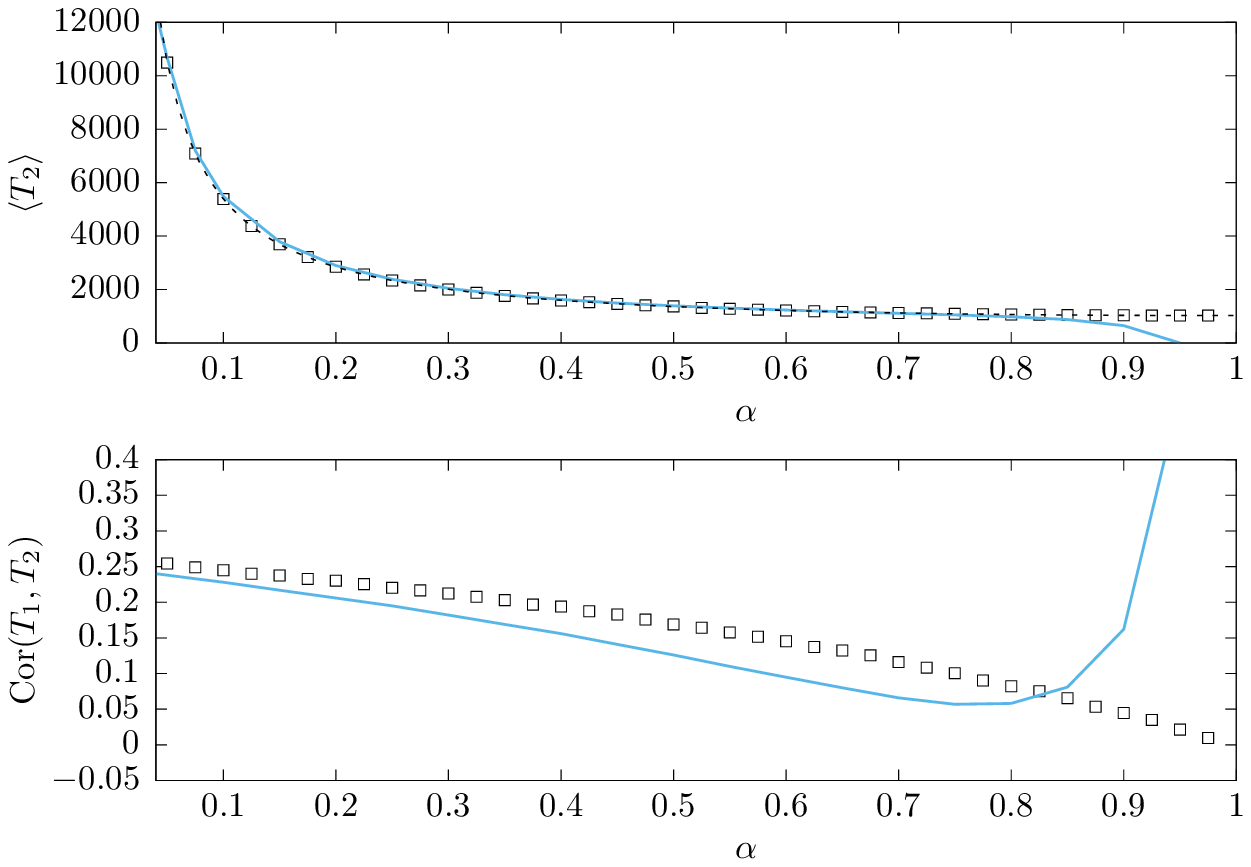}
	\caption{Comparison for the (a) mean recurrence time after first hitting, and (b) correlation coefficient between first hitting and first recurrence times as a function of the memory parameter $\alpha$, between results obtained from Monte Carlo simulations (symbols) averaged over $10^5$ realizations, and the analytical approximation reported in Section \ref{approx} (solid line). Different Results are shown for the case without spatial constraints and an exponential waiting time distribution $\varphi(t)=\tau^{-1}e^{-t/ \tau}$ with $\tau=1$. The dotted line represents the heuristic approximation $\langle T_2 \rangle = \tau N / \alpha (2-\alpha)$.}
	\label{fig2}
\end{figure}

In Figure 2 we see that our analytical approximations for $\langle T_2 \rangle$ and $\text{Cor} (T_1,T_2)$ (solid lines) fits rather well the results for $\alpha$ small but the approximation fails clearly as long as we approach the limit without memory, $\alpha=1$. Surprisingly, we find that a simple heuristic approximation is able to fit the behavior for $\langle T_2 \rangle$ accurately (dotted lines in Figure 2). Such approximation follows from assuming that the probability of reaching the target at each particular jump is constant and equals $\alpha /N + (1-\alpha) \alpha / N = \alpha (2-\alpha)/N$. Here, the two terms in the sum represent the contribution from the \textit{normal} and \textit{memory} modes, respectively. For the latter we are assuming that the individual revisits the target with probability $(1-\alpha)/\langle m \rangle $, where $\langle m \rangle$ is the mean number of jumps done previously, $N/ \alpha$ (as discussed in Section \ref{theory}). The rate at which the target is found will be then $\tau$ times the inverse of the overall probability, leading to the estimation $\langle T_2 \rangle = \tau N / \alpha (2-\alpha)$. The agreement found in Figure 2 between this approximation (dotted line) and the numerical results is clear. However, this kind of approximation cannot be extended to cross moments (e.g., $\langle T_1 T_2 \rangle$) and so it is not helpful to obtain an estimation of the correlation coefficient, as we pursue here.

Regarding the correlation coefficient, Figure 2 shows that the tendency predicted by our analytical approach is approximately correct (except, again, in the limit $\alpha \rightarrow 1$), but the accumulated error in the estimation of $\langle T_2 \rangle$, $\langle T_2^2 \rangle$ and $\langle T_1 T_2 \rangle$ makes that the quantitative agreement is not completely satisfactory. In any case, this plot confirms the main idea of the present work, which is the fact that consecutive hitting times become positively correlated as a consequence of the preferential relocation rule.

\begin{figure}
	\includegraphics[scale=1.0]{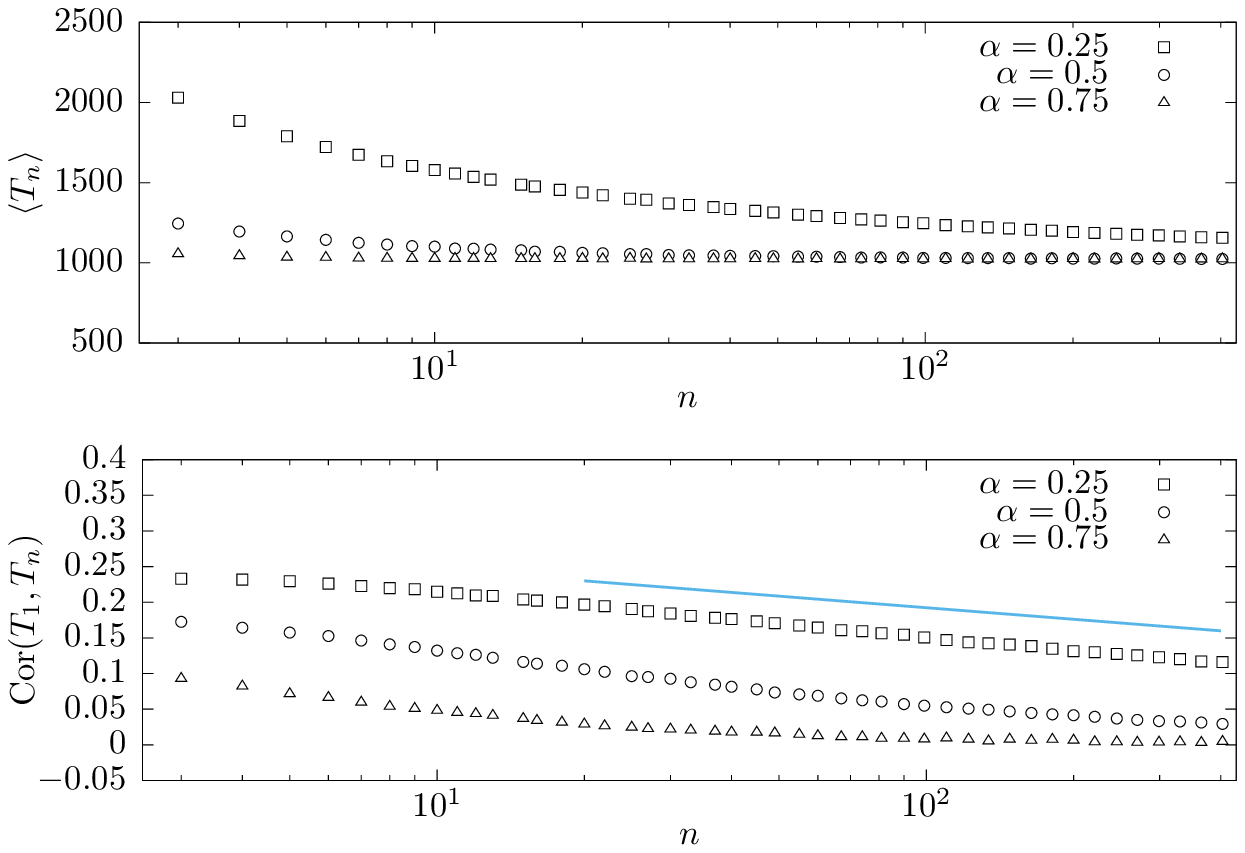}
	\caption{Results for the (a) mean recurrence time after first hitting and (b) correlation coefficient between first-passage and subsequent recurrence times as a function of $n$, obtained from Monte Carlo simulations averaged over $10^5$ realizations. Different values of the memory parameter $\alpha$ are reported (see legends). The solid line in the lower panel is just a visual cue to emphasize the logarithmic character of the decay.}
	\label{fig3}
\end{figure}

Most of the conclusions above for $n=2$ can be extended to subsequent recurrence times ($n>2$), as reported in Figure 3. There we observe that the mean value of $T_n$ becomes progressively reduced as a function of $n$ due to the accumulated effect of memory, and also it can be checked that its behavior as a function of the memory parameter $\alpha$ is qualitatively the same as for $T_2$ (not shown). Furthermore, Figure 3b yields a very interesting result, as is the fact that correlations between first-passage and subsequent recurrence times $T_n$ show an ultra-slow (logarithmic, as for the MSD) relaxation to zero as a function of $n$. This tells us that the signature of strong memory induced by the preferential relocation rule does not only emerge at the level of dispersal (as was already known from previous works \cite{boyer14,boyer17}) but also on the recurrence statistics.

\begin{figure}
	\includegraphics[scale=1.0]{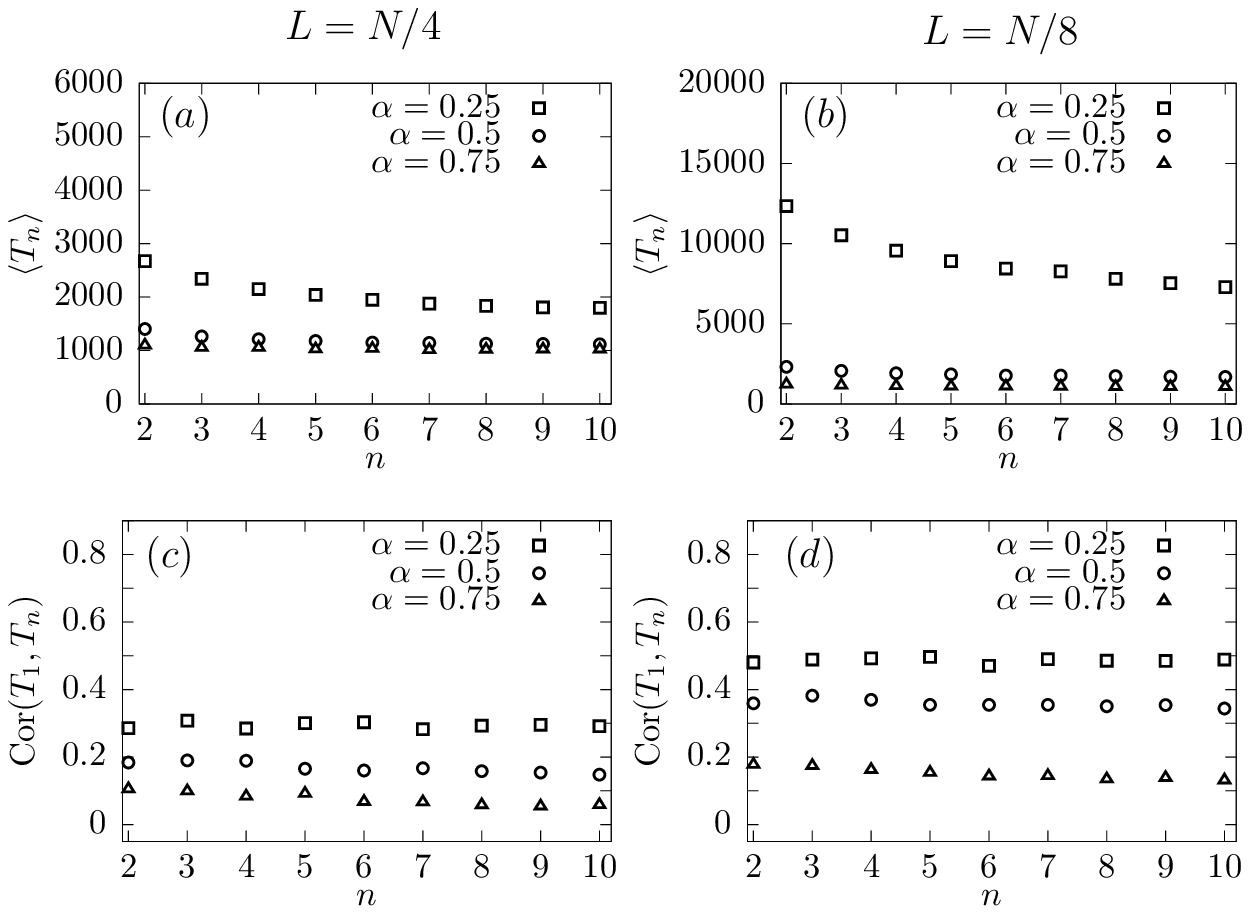}
	\caption{Results for the (a,b) mean recurrence time after first-hitting and (c,d) correlation coefficient between first-passage and subsequent recurrence times as a function of $n$, obtained from Monte Carlo simulations, for the case where the \textit{normal} mode imposes a constraint given by a maximum jump length of size $L$. Two values of $L$ and different values of the memory parameter $\alpha$ are reported (see legends) for the sake of completeness.}
	\label{fig4}
\end{figure}

Finally, to complete the numerical analysis we provide results for the case where individuals are not allowed to jump freely from any site to another, but only short jumps are to be expected.
So that, we reformulate the dispersal process by assuming that the individual can only do jumps up to a maximum distance $L$ (this rule only applies to the \textit{normal} mode, while the \textit{memory} mode is kept unchanged and without spatial constraints). The results for $L=N/4$ and $L=N/8$ are provided in Figure 4, where the initial position is chosen at random at each realization of the Monte Carlo simulation. 

Note that the spatial constraint increases considerably the values of the first-passage and recurrence times (since now further sites become increasingly difficult to be reached due to the ultra-slow dispersal properties of the model). So, low values of $L$ become computationally very costly. Apart from that, we observe that the spatial constraint does not modify qualitatively the picture found in the Figures 2 and 3. In particular, positive correlations between first-passage and recurrence times are still present and actually become inforced.

\section{Conclusions}
\label{conclusions}

Summarizing the ideas reported in the work, we have confirmed that one of the essential signatures of random walks with memory (in this case, we have just focused on the model with preferential relocation to visited places) is the existence of correlations between first-passage and subsequent recurrence times through a site or, equivalently, between consecutive recurrence times. Furthermore, we have been able to characterize such correlations not only numerically but also through an approximated analytical study, at least for the case $n=2$.

Anyway, the most interesting result we obtain is probably (FIgure 3) that the decay of these correlations with time for the case of walks with preferential relocation exhibits an ultra-slow (logarithmic) behavior, in accordance with the dispersal properties of the model which were already known from previous works \cite{boyer14,boyer14b}. Since memory effects persist in the model for arbitrarily long times, such long-range dynamics is also present in the hitting and recurrence statistics. This suggests that such correlations capture adequately the memory dynamics in the model and so can be used as a proxy to identify what kind of memory rules govern the process. This could be of great interest, for example, in the analysis of trajectories of real organisms as a method to understand how memory has been employed during the process.

\section*{Acknowledgements} This research has been supported by the Spanish government through Grants No. CGL2016-78156-C2-2-R and FIS2015-72434-EXP.

\section{APPENDIX. Derivation of the moments of $T_2$}
\label{appendix}

We start from the definition of the coefficients $B_{m,m_1}$, given implicitly in (\ref{n22}), and express them in a more convenient form:
\begin{equation}
B_{m,m_1} = \left[ \gamma (m_1) \right] ^m \frac{\left( 1+ \frac{1}{m_1-\beta} \right) \ldots \left( 1+ \frac{n-1}{m_1-\beta} \right) }{\left( 1+ \frac{1}{m_1} \right) \ldots \left( 1+ \frac{n-1}{m_1} \right)},
\end{equation}
where we have defined $\beta \equiv (1-\alpha)/(1-\alpha/N)$ and $\gamma (m_1) \equiv (1-\alpha /N)(1-\beta/m_1)$. In this way it is clear that we can propose an expansion for $m_1 \gg 1$ in the form 
\begin{eqnarray}
B_{m,m_1} &=& \left[ \gamma (m_1) \right] ^m \left( 1+\sum_{i=1}^{m-1} \frac{v_i(m)}{(m_1-\beta)^i} \right) \nonumber\\
&\times &\left( 1-\sum_{i=1}^{m-1} \frac{w_i(m)}{(m_1)^i} \right),
\end{eqnarray}
with the first coefficients $v_i$, $w_i$ given by
\begin{eqnarray}
\nonumber v_1(m) &=& w_1(m) = \frac{m (m-1)}{2} \\
\nonumber v_2(m) &=& \frac{m (m-1) (m-2) (3m-1)}{72} \\
\nonumber w_2(m) &=& v_2(m)-v_1^{2}(m) \\ 
\ldots
\end{eqnarray}
Leading this expansion up to second order (this is, up to powers of order $m_1^{-2}$) and summing over all values of $m=0$ we obtain, after some lengthy algebra,
\begin{eqnarray}
\nonumber \sum_{m=0}^{\infty} B_{m,m_1} &=& \frac{1}{1-\gamma(m_1)} + \frac{ (1-\alpha) \gamma (m_1)}{m_1^2 (1-\gamma(m_1))^3} - \\
&-& \frac{\beta \left( 1+ \gamma (m_1) \right) }{m_1^3 (1- \gamma(m_1))^4} + \frac{\beta^2 (1+2 \gamma (m_1)) }{m_1^4 (1-\gamma (m_1))^5}\nonumber \\ 
&+& \ldots 
\label{bm}
\end{eqnarray}
In order to apply (\ref{t2}) we use the fact that in the limit of large media size ($N \gg 1$, $m_1 \gg 1$) the sum over $m_1$ can be adequately approximated by an integral
\begin{eqnarray}
\langle T_2 \rangle &=& \sum_{m_{1}=0}^{\infty} \frac{\alpha \tau}{N} \left( 1- \frac{\alpha}{N} \right) ^{m_1-1} \left( \sum_{m=0}^{\infty} B_{m,m_1} \right) \nonumber \\
&\approx& \frac{\alpha \tau}{N} \int_{0}^{\infty}  e^{-\frac{\alpha}{N} u} \left( \sum_{m=0}^{\infty} B_{m,u} \right) \text{d} u.
\label{t2aprox}
\end{eqnarray}
So, by introducing (\ref{bm}) into (\ref{t2aprox}) one finally obtains an approximated expression for $\langle T_2 \rangle$. This result can be expressed as a combination of algebraic and exponential integral functions (the resulting expression is too long to be reproduced here).

Similarly, the second order moment of $\langle T_2 \rangle$ can be approximated using exactly the same procedure and leading to
\begin{equation}
\langle T_2^2 \rangle \approx \frac{2 \alpha \tau^2}{N} \int_{0}^{\infty}  e^{-\frac{\alpha}{N} u} \left( \sum_{m=0}^{\infty} (1+m) B_{m,u} \right) \text{d} u.
\label{t2aproxb}
\end{equation}
and the same for the cross moment,
\begin{equation}
\langle T_1 T_2 \rangle \approx \frac{\alpha \tau^2}{N} \int_{0}^{\infty}  u e^{-\frac{\alpha}{N} u} \left( \sum_{m=0}^{\infty} B_{m,u} \right) \text{d} u
\end{equation}
so the value of the correlation coefficient between $T_1$ and $T_2$, as defined in (\ref{cor}), can be estimated.

\newpage


\begin{thebibliography}{99}
	
	\bibitem{papo17} Papo D, Go\~{n}i J and J.M. Bold\'{u} 2017 On the relation of dynamics and structure in brain networks \textit{Chaos} 27, 047201.
	\bibitem{chialvo10} Chialvo D R 2010 Emergent complex neural dynamics \textit{Nature Phys.} 6, 744.
	\bibitem{codling08} Codling E A 2008 Random walk models in biology \textit{J. Roy. Soc. Interface} 5, 813.
	\bibitem{mendez14} M\'{e}ndez V, Campos D and Bartumeus F 2014 \textit{Stochastic Foundations in Movement Ecology}. (Berlin: Springer-Verlag).
	\bibitem{bartumeus16} Bartumeus F et. al. 2016  Foraging success under uncertainty: search tradeoffs and optimal space use \textit{Ecol. Lett.} 19, 1299.	
	\bibitem{holmes93} Holmes E E 1993 Are Diffusion Models too Simple? A Comparison with Telegraph Models of Invasion \textit{Am. Nat.} 142, 779.
	\bibitem{selmeczi05} Selmeczi D, Mosler S, Hagedom P H, Larsen N B and Flyvbjerg H 2005 Cell motility as persistent random motion: theories from experiments \textit{Biophys. J.} 89, 912.
	\bibitem{shlesinger18} Shlesinger M F 2018 Origins and applications of the Montroll-Weiss continuous time random walk. \textit{Eur. Phys. J. B} 91:40.
	\bibitem{viswanathan96} Viswanathan G M et. al. 1996 L\'{e}vy flight search patterns of wandering albatrosses \textit{Nature} 381, 413
	\bibitem{ariel15} Ariel G et. al. 2015 Swarming bacteria migrate by L\'{e}vy Walk \textit{Nature Comm.} 6:8396.
	\bibitem{freund92} Freund H and Grassberger P 1992 The Red Queen's walk \textit{Physica A} 190, 218-237.
	\bibitem{schutz04} Sch\"{u}tz G M and Trimper S 2004 Elephants can always remember: Exact long-range memory effects in a non-Markovian random walk \textit{Phys. Rev. E} 70, 045101(R).
	\bibitem{boyer14} Boyer D and Sol\'{i}s-Salas C 2014 Random Walks with Preferential Relocations to Places Visited in the Past and their Application to Biology \textit{Phys. Rev. Lett.} 112, 240601.
	\bibitem{falcon17} Falc\'{o}n-Cort\'{e}s A, Boyer D, Giuggioli L and Majumdar S N 2017 Localization Transition Induced by Learning in Random Searches \textit{Phys. Rev. Lett.} 119, 140603.
	\bibitem{abramson14} Abramson G, Kuperman M N, Morales J M and Miller J C 2014 Space use by foragers consuming renewable resources \textit{Eur. Phys. J. B} 87:100.
	\bibitem{kazimierski15} Kazimierski L D, Abramson G and Kuperman M 2015 Random-walk model to study cycles emerging from the exploration-exploitation trade-off \textit{Phys. Rev. E} 91, 012124.
	\bibitem{guerin16} Gu\'{e}rin T, Levernier N, B\'{e}nichou O and Voituriez R 2016 Mean first-passage times of non-Markovian random walkers in confinement \textit{Nature} 534, 356.
	\bibitem{campos16} Campos D, Bartumeus F and M\'{e}ndez V 2016 Nonstationary dynamics of encounters: Mean valuable territory covered by a random searcher \textit{Phys. Rev. E} 96, 032111.
	\bibitem{kearney16} Kearney M J and Martin R J 2016 First passage properties of a generalized P\'{o}lya urn \textit{J. Stat. Mech: Theor. Exp.} 123407.
	\bibitem{boyer14b} Boyer D and Romo-Cruz J C R 2014 Solvable random-walk model with memory and its relations with Markovian models of anomalous diffusion \textit{Phys. Rev. E} 90, 042136.
	\bibitem{boyer17} Boyer D, Evans M R and Majumdar S N 2017 Long time scaling behaviour for diffusion with resetting and memory \textit{J. Stat. Mech: Theor. Exp.} 023208.
	\bibitem{MeIo} M\'{e}ndez V, Iomin A, Horsthemke W and  Campos D 2017 Langevin dynamics for ramified structures \textit{J. Stat. Mech: Theor and Exp.} 063205.
	\bibitem{iomin16} Iomin A and M\'{e}ndez V 2016 Does ultra-slow diffusion survive in a three dimensional	cylindrical comb? \textit{Chaos, Solitons \& Fractals} 82, 142. 
	\bibitem{condamin05} Condamin S, B\'{e}nichou O and Moreau M 2005 First-Passage Times for Random Walks in Bounded Domains \textit{Phys. Rev. Lett.} 95, 260601.
	\bibitem{redner01} Redner S 2001 \textit{A Guide to First-Passage Processes} (Cambridge: Cambridge Univ. Press).
	
\end{thebibliography}
\end{document}